\documentclass[twoside,twocolumn,9pt]{article}
\usepackage{extsizes}
\usepackage[super,sort&compress,comma]{natbib} 
\usepackage[left=1.5cm, right=1.5cm, top=1.785cm, bottom=2.0cm]{geometry}
\usepackage{balance}

\usepackage{sectsty}
\allsectionsfont{\scshape}

\usepackage{graphicx} 
\usepackage{lastpage}
\usepackage{float}
\usepackage{fancyhdr}
\usepackage{fnpos}
\usepackage[english]{babel}
\usepackage{array}
\usepackage[T1]{fontenc}
\usepackage[usenames,dvipsnames]{xcolor}
\usepackage{setspace}
\usepackage{hyperref}

\usepackage{upgreek}
\usepackage{xspace}
\usepackage{placeins}
\usepackage{amsmath}

\newcommand{\ds}{\ensuremath{d_\mathrm{S}}\xspace}
\newcommand{\dl}{\ensuremath{d_\mathrm{L}}\xspace}
\newcommand{\Vs}{\ensuremath{V_\mathrm{S}}\xspace}
\newcommand{\Vl}{\ensuremath{V_\mathrm{L}}\xspace}
\newcommand{\Vf}{\ensuremath{V_\mathrm{f}}\xspace}
\newcommand{\NS}{\ensuremath{N_\mathrm{S}}\xspace}
\newcommand{\NL}{\ensuremath{N_\mathrm{L}}\xspace}
\newcommand{\phic}{\ensuremath{\phi_\mathrm{c}}\xspace}
\newcommand{\tds}{\tilde{\ds}}
\newcommand{\tdl}{\tilde{\dl}}
\newcommand{\tc}{\ensuremath{t_\mathrm{c}}\xspace}
\newcommand{\dt}{\ensuremath{\Delta t}\xspace}
\newcommand{\veta}{\ensuremath{v_\eta}\xspace}
\newcommand{\etae}{\ensuremath{\eta_\mathrm{exp}}\xspace}
\newcommand{\etaf}{\ensuremath{\eta_\mathrm{f}}\xspace}
\newcommand{\etar}{\ensuremath{\eta_\mathrm{r}}\xspace}

\newcommand{\um}{\xspace\ensuremath{{\rm \mu m}}\xspace}
\newcommand{\mm}{\xspace\ensuremath{{\rm mm}}\xspace}
\newcommand{\ms}{\xspace\ensuremath{{\rm ms}}\xspace}
\newcommand{\mms}{\xspace\ensuremath{{\rm mm/s}}\xspace}

\begin{document}

\twocolumn[
  \begin{@twocolumnfalse}
\vspace{3cm}
% \sffamily

\begin{center}

%     Title
    \noindent\huge{\textbf{\textsc{Droplet detachment and pinch-off of bidisperse\\ particulate suspensions}}} \\
    \vspace{1cm}

%     Authors
    \noindent\large{Virgile Thi\'evenaz,\textit{$^{a}$} Sreeram Rajesh,\textit{$^{a}$} and Alban Sauret\textit{$^{a}$}}$^{\ast}$ \\%Author names go here instead of "Full name", etc.

    \vspace{5mm}
%     Date
    \noindent\large{\today} \\%Author names go here instead of "Full name", etc.

    \vspace{1cm}
    \textbf{\textsc{Abstract}}
    \vspace{2mm}

\end{center}

\noindent\normalsize{
    When a droplet is generated, the ligament connecting the drop to the nozzle thins down and 
    eventually pinches off. 
    Adding solid particles to the liquid phase leads to a more complex dynamic,
    notably by increasing the shear viscosity.
    Moreover, it introduces an additional length scale to the system, the diameter of the particles,
    which eventually becomes comparable to the diameter of the ligament. 
    In this paper, we experimentally investigate the thinning and pinch-off of drops of suspensions with two
    different sizes of particles.
    We characterize the thinning for different particle size ratios and different proportions of small particles.
    Long before the pinch-off, the thinning rate is that of an equivalent liquid whose viscosity is that
    of the suspension.
    Later, when the ligament thickness approaches the size of the large particles, 
    the thinning accelerates and leads to an early pinch-off.
    We explain how the bidisperse particle size distribution lowers the viscosity 
    by making the packing more efficient, which speeds up the thinning.
This result can be used to predict the dynamics of droplet formation with bidisperse suspensions.} \\

 \end{@twocolumnfalse} \vspace{0.6cm}

  ]

\makeatletter
{\renewcommand*{\@makefnmark}{}
\footnotetext{\textit{$^{a}$~Department of Mechanical Engineering, University of California, Santa Barbara, California 93106, USA}}
\footnotetext{\textit{$^{*}$ asauret@ucsb.edu}}
\makeatother
%   \footnotetext[~]{\textit{$^{a}$~Department of Mechanical Engineering, University of California, Santa Barbara, California 93106, USA; E-mail: asauret@ucsb.edu}}

%%%MAIN TEXT%%%%
%%%%%%%%%%%%%%%%%%%%%%%%%%%%%%%%%%%%%%%%%%%%%%%%%%%%%%%%%%%%%%%
%%%%%%%%%%%%%%%%%%%  Introduction   %%%%%%%%%%%%%%%%%%%%%%%%%%
%%%%%%%%%%%%%%%%%%%%%%%%%%%%%%%%%%%%%%%%%%%%%%%%%%%%%%%%%%%%%%%

\section{Introduction}

The generation of droplets of suspension, \textit{i.e.}, a fluid containing a solid dispersed phase, 
is present in many printing processes.\cite{studart2016,tan2020} 
For example, bio-printing frequently requires the inclusion of cells or biomaterials in a liquid matrix.
\cite{nakamura2005,gudapati2016}
The generation of suspension droplets is related to the printability of the fluid, 
which is influenced by the nature of the liquid\cite{thievenaz2021} 
and that of the particles.\cite{ketel2019,chan2020}
For a suspension, the formation of drops is first controlled by the rheological behavior resulting from the presence of the particles. 
At first order, the viscosity of the suspension increases with the solid volume fraction $\phi$.\cite{stickel2005,denn2014,guazzelli2018}
However, the presence of solid particles also modifies the pinch-off dynamics,\cite{furbank2006} 
in particular when their size becomes comparable to the length scale of the flow.
This effect is particularly important during the pinch-off of a drop since the diameter of the ligament becomes vanishingly small.\cite{eggers1994}
Such a deviation from a continuous-medium behavior has also been observed during the deposition of thin films of suspensions on substrates
\cite{gans2019,palma2019,jeong2020} and in the atomization of suspension sheets.\cite{addo-yobo2011,raux2020}

Various manufacturing processes involve the generation of drops at different scales.\cite{teh2008,vaezi2013}
Many studies have investigated the pinch-off dynamics of homogeneous Newtonian liquids extruded from a nozzle to describe and optimize these processes.
\cite{eggers2008} 
The formation of drops directly at the nozzle is observed in the dripping regime when the extrusion speed of the fluid is low enough and is accompanied by a localized break-up.\cite{clanet1999} 
Experiments performed with homogeneous Newtonian fluids have shown that the thickness of the liquid neck,
which connects the drop to the nozzle, vanishes in a finite time $\ensuremath{t_\mathrm{c}}\xspace$.\cite{eggers2008}
Near this pinch-off singularity, the thinning follows a self-similar dynamic.
The relevant length scale is no longer the nozzle diameter but the thickness of the liquid neck $h(t)$ at its narrowest point,
and the relevant time scale is the time to the pinch-off $\tc-t$.
The mechanisms acting on the liquid neck, which are captured by the fluid viscosity $\eta$, its surface tension $\gamma$,
and the inertia through its density $\rho$, can be summed up as the Ohnesorge number, ${\rm Oh}={\eta}/{\sqrt{\rho \gamma h}}$. This dimensionless number represents the ratio of viscous to inertial forces in a capillary flow.
In the inviscid limit of small ${\rm Oh}$, dimensional analysis shows that 
$h(t) \propto \left( \gamma / \rho \right)^{1/3} \left(\tc-t \right)^{2/3}$,\cite{eggers1993,day1998} 
whereas in the viscous limit of large ${\rm Oh}$, 
the neck diameter evolves as $h(t) \propto (\gamma/\eta)\,\left(\tc-t \right)$.\cite{papageorgiou1995} 

The presence of solid particles dispersed in the fluid complexifies the problem
since near the pinch-off the diameter of the neck becomes comparable to that of the particles.
Various studies on the dynamics of jets and drops of particulate suspensions have shown that the dripping/jetting transition occurs at lower flow rates for a particulate suspension than for an equivalent homogeneous liquid
with the same viscosity.\cite{furbank2006,chateau2019}
Past experiments have been performed with suspensions of monodisperse particles.
Hence, the flow was characterized by two length scales: the particle diameter $d$ and the thickness $h(t)$ of the jet (or neck)
at its narrowest point.
These experiments have revealed two regimes: a continuous regime at early-time and a regime presenting discrete effects brought by the discontinuous nature of the solid phase near the pinch-off.\cite{furbank2006,bonnoit2012,bertrand2012,miskin2012,vandeen2013}
Similar studies on stretched ligaments of suspensions led to the same conclusions.
\cite{mcIlroy2014,lindner2015,moon2015,mathues2015,chateau2018} 
In these configurations, it was shown that as long as $h(t) \gg d$, the suspension can be considered as an effective homogeneous Newtonian fluid
whose viscosity is independent of the size of the particles.\cite{guazzelli2018}
This continuous description fails near the final pinch-off, and the thinning accelerates.
In the case of viscous interstitial fluids and monodisperse suspensions, it has recently been shown that the transition to the discrete effects regime occurs
at the critical neck thickness $h^{\star} \sim d \left(\phi_{\mathrm{c}}-\phi\right)^{-1 / 3}$, 
where $\phic$ denotes the maximum packing fraction.\cite{chateau2018}
Bonnoit \textit{et al.}\cite{bonnoit2012} have also reported that this discrete effects regime is followed 
for a short period by a regime only controlled by the interstitial liquid,
identical to the pinch-off of a drop of pure interstitial liquid.
Moreover, for suspensions, the pinch-off is very localized, whereas for a viscous fluid having the same viscosity,
a long and thin filament is observed until the drop detaches.\cite{furbank2006,bonnoit2012}

The rheology of particulate suspensions has been extensively characterized for monodisperse, spherical, neutrally buoyant, 
and non-Brownian particles dispersed in a Newtonian liquid.\cite{guazzelli2018}
At small particle Reynolds numbers, the suspension exhibits a Newtonian rheology. 
The effective shear viscosity $\eta(\phi)$ of the suspension depends on the interstitial fluid viscosity \etaf
and on the particle volume fraction $\phi$ but not on the particle diameter $d$.
Different empirical correlation for $\eta(\phi)$ have been proposed.
\cite{zarraga2000,boyer2011,guazzelli2018}
For instance, the Maron-Pierce model captures experimental measurements reasonably well,
while retaining a simple expression:\cite{maron1956}
\begin{equation}
    \eta(\phi) = \etaf (1 - \phi/\phic)^{-2},
    \label{eq:maronpierce}
\end{equation}
where \phic is the maximum volume fraction for a suspension of
monodisperse spherical particles.
Its value depends on the inter-particles friction coefficient and is typically in the range $0.56 < \phic < 0.64$.\cite{guazzelli2018}
The quantity $1 - \phi/\phic$ roughly describes the volume in which each particle is free to move without hindrance from its neighbors.
When $\phi \to \phic$, this leeway vanishes, and the suspension jams.
Such rheological approaches have been shown to capture the early thinning
of monodisperse suspension thread.\cite{bonnoit2012,chateau2018}

The studies mentioned above all considered monodisperse suspensions where the particles are described by a single length scale.
However, in practical applications, the dispersed phase often features a broad particle size distribution,
and such ideal descriptions fail to capture the reality.
The viscosity of bidisperse suspensions is more challenging to describe.
\cite{shapiro1992,chang1994,gondret1997,gamonpilas2016,madraki2017enhancing,pednekar2018,madraki2018transition}
The main observation is that bidisperse suspensions display a lower viscosity than monodisperse ones for the same volume fraction $\phi$.
This observation is correlated to the increase in the maximum volume fraction $\phic$.\cite{gondret1997,gamonpilas2016}
In order to describe the rheology of bidisperse suspensions, two additional parameters are required:
the ratio of large to small particle diameters, $\delta=\dl/\ds$,
and the fraction of the solid volume occupied by the small particles,
$\xi=\Vs/(\Vs+\Vl)$.\cite{pednekar2018}

The capillary dynamics of bidisperse suspensions remains elusive. 
One may intuitively expect that the macroscopic viscosity will be reduced. 
However, the free surface may cause the reorganization, filtration, or sorting of the particles of different sizes, as recently observed for dip-coating.
\cite{sauret2019,dincau2019,dincau2020}
The pinch-off dynamics and the droplet formation have only been characterized 
for monodisperse suspensions, and our understanding and modeling of the printing of polydisperse suspensions
remains limited.
In particular, for a bimodal distribution of particles, the introduction of the new parameters $\delta$ and $\xi$ makes uncertain whether the models for the pinch-off of
monodisperse suspensions hold in the bidisperse case.

To characterize this configuration, we consider in this study a model bidisperse suspension 
and investigate experimentally the pinch-off and detachment of droplets.
In particular, we report the thinning dynamics of the liquid neck between the drop and the nozzle.
We start with a presentation of our experimental approach,
and we report qualitative observations between different compositions of particulate suspensions.
Then, we quantify the time evolution of the neck thickness in the different regimes before pinch-off.
Finally, we discuss these results in light of the rheology of bidisperse suspensions.

%%%%%%%%%%%%%%%%%%%%%%%%%%%%%%%%%%%%%%%%%%%%%%%%%%%%%%%%%%%%%%%
%%%%%%%%%%%%%%%%%  Experimental Methods   %%%%%%%%%%%%%%%%%%%%%
%%%%%%%%%%%%%%%%%%%%%%%%%%%%%%%%%%%%%%%%%%%%%%%%%%%%%%%%%%%%%%%

\begin{figure*}[p]
    \centering
    \includegraphics[width=\textwidth]{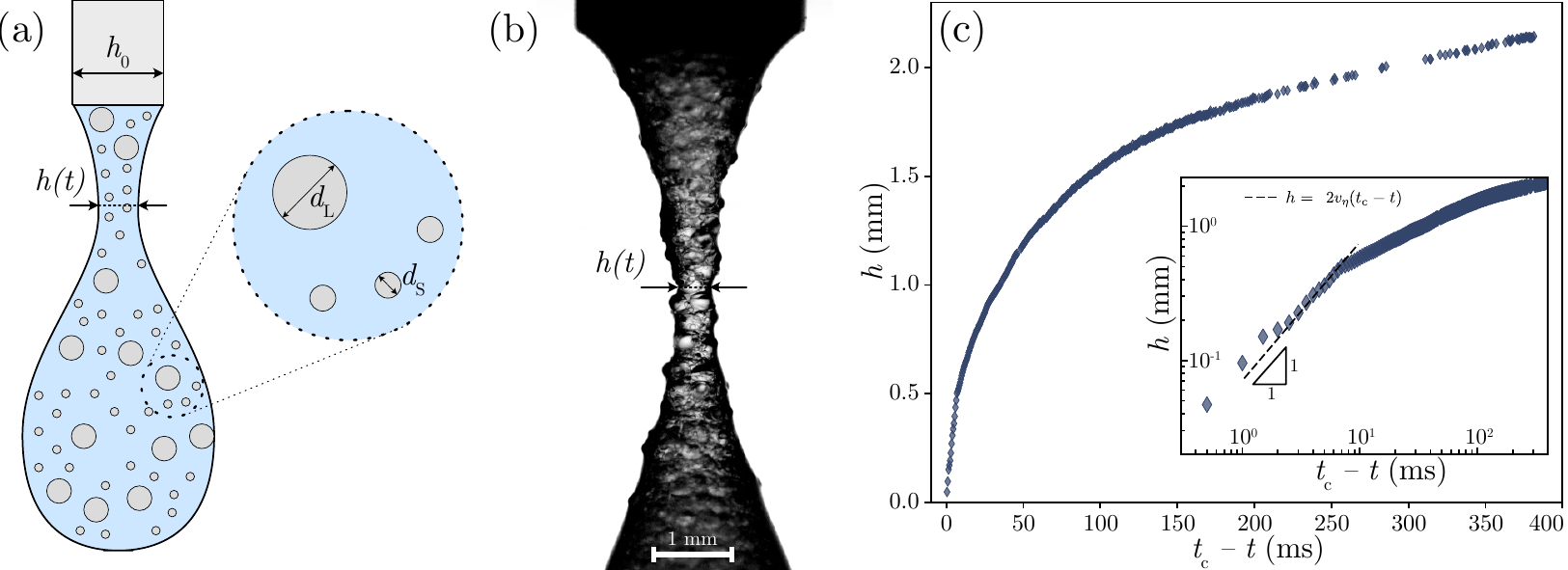}
    \caption{(a) Schematic of the experimental method.
            (b) Picture of the neck of a bidisperse suspension drop containing 
            $\phi=40\%$ of particles distributed as 
            $\xi = 20\%$ of small particles ($\ds=80\,\mu{\rm m}$) 
            and $1-\xi = 80\%$ of large particles ($\dl = 250\,\mu{\rm m}$).
            (c) Thinning dynamics of the neck for the same bidisperse suspension. The inset shows the dynamics in log scale.
            The dashed line represents the linear self-similar regimes for capillary-viscous
            thinning.
        }
    \label{fig:quali}
\end{figure*}

\begin{figure*}[p]
    \centering
    \includegraphics[width=\textwidth]{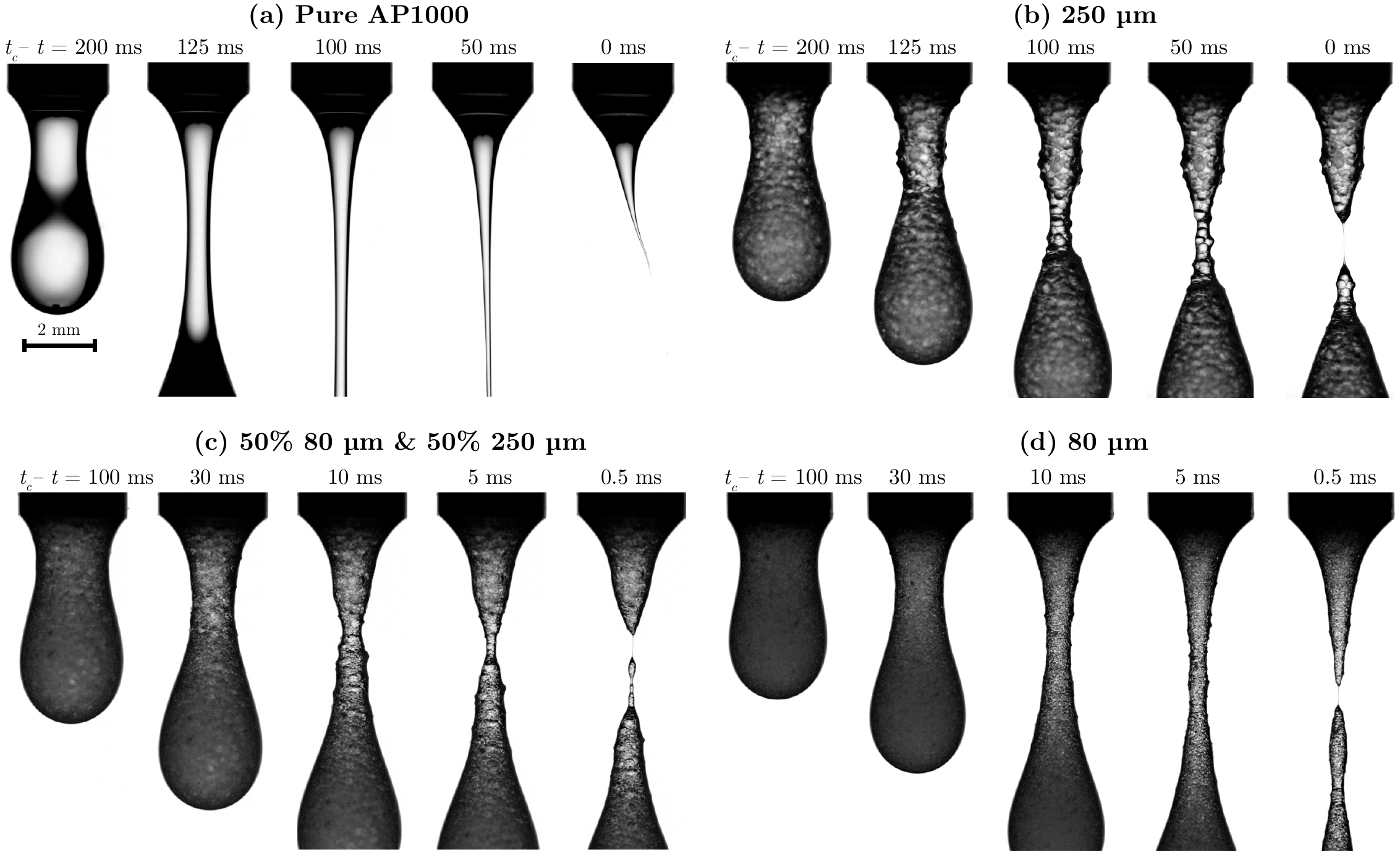}
    \caption{Example of droplet detachment for (a) AP1000 silicone oil (no particles)
        and for suspensions containing $\phi = 40\%$ of particles distributed as follows:
        (b) $250\,\mu{\rm m}$ particles only, corresponding to $\xi = 0$;
        (c) half of $80\,\mu{\rm m}$ and half of $250\,\mu{\rm m}$ particles, corresponding to $\xi = 0.5$;
        (d) $80\,\mu{\rm m}$ particles only, \textit{i.e.}, $\xi = 1$. The corresponding movies are available in Supplemental Materials.
    }
    \label{fig:timeline}
\end{figure*}

\begin{figure*}[h]
    \centering
    \includegraphics[width=0.99\textwidth]{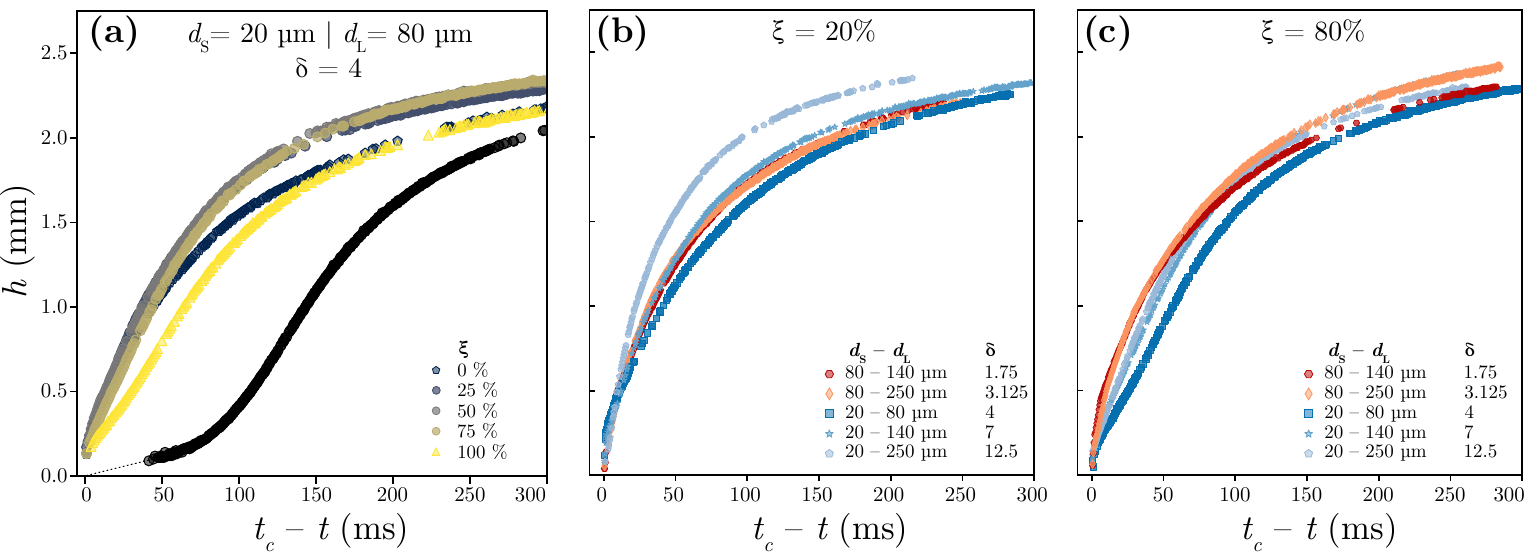}
    \caption{Thinning dynamics for the pinch-off of different bidisperse 
            suspensions droplets. The total volume fraction remains constant and equal to $\phi=40\%$.
            (a) Different values of the small particle fraction $\xi$ for a bidisperse suspension composed
            of particles of diameters \ds = $20\,\um$ and \dl = $80\,\um$.
            The black circles represent the experiment with the AP1000 silicone oil.
            (b) For a larger fraction of big particles ($\xi = 20\%$) and different bidisperse suspensions
            of small and large particle diameters of \ds and \dl, respectively.
            (c) Situation with a larger fraction of small particles ($\xi = 80\%$) 
            for different bidisperse systems.
            }
    \label{fig:dynamics}
\end{figure*}

\section{Experimental Methods}

The suspensions consist in different batches of spherical 
polystyrene particles (Dynoseeds from Microbeads),
with density $\rho = 1054 \pm 4 \,{\rm kg.m^{-3}}$ and diameters 20, 80, 140, and 250 \um.
The particles are dispersed in silicone oil (AP100, Polyphenyl-methylsiloxane from Sigma-Aldrich) 
of dynamic viscosity $\etaf = 120\,{\rm mPa.s}$ and density 
$\rho=1058\,{\rm kg.m^{-3}}$ at $20^\circ \text C$. 
The interstitial liquid was chosen to match the density of the particles 
to limit sedimentation effects over the duration of the experiments. 
The silicone oil perfectly wets the particles, although it was shown that the contact angle between the particles and the interstitial liquid does not affect the pinch-off dynamics.\cite{chateau2018}
To separate the influence of the effective viscosity from the discrete effects of the particles, 
we also compare the behavior of the suspensions with that of a Newtonian fluid having similar macroscopic properties as the suspension.\cite{bonnoit2012,raux2020} 
Here, we use a more viscous silicone oil (AP1000, $\eta_\text{AP1000}=1.28\,{\rm Pa.s}$ at $20^\circ \text C$), 
whose viscosity is close to the viscosity of a monodisperse suspension of volume fraction $\phi=40\%$.

The particle volume fraction is $\phi=({\Vs+\Vl})/({\Vs+\Vl+\Vf})$,
with \Vs and \Vl the volume of small and large particles, respectively, and \Vf the volume of interstitial liquid.
Since we focus on the role of the bidisperse distribution for a given volume fraction, 
we keep it constant at $\phi=40\%$.
To characterize the role of the size distribution,
we vary the size ratio of large to small particle diameters $\delta=\dl/\ds$
and the fraction of the total solid volume occupied by the small particles $\xi=\Vs/(\Vs+\Vl)$. 

The experiments consist of slowly extruding the suspension from a syringe with a nozzle
of outer diameter $h_0=2.75\mm$ [figure~\ref{fig:quali}(a)]. 
The formation of the drop and the pinch-off are recorded with a high-speed camera 
(Phantom VEO710) at 2,000 fps. 
The camera is equipped with a macro lens (Nikon 200mm f/4 AI-s Micro-NIKKOR)
and a microscope lens (Mitutoyo X2) so that the typical resolution in our measurement is 10\um.
The experiments are backlit with a LED panel (Phlox) to clearly see the contour
of the drop and the suspension thread [figure~\ref{fig:quali}(b)].
This contour is detected with thresholding methods using the software ImageJ.
Then, a custom Python routine enables us to extract the minimum thickness of the ligament, 
$h(t)$, at each time step.
Figure~\ref{fig:quali}(c) reports an example of the time evolution of the minimum thickness $h$ \ in the case:
$\ds = 80\um$, $\dl = 250\um$, $\phi = 40\%$ and $\xi=20\%$.
For each suspension, experiments were repeated five times and led to reproducible results.

%%%%%%%%%%%%%%%%%%%%%%%%%%%%%%%%%%%%%%%%%%%%%%%%%%%%%%%%%%%%%%%
%%%%%%%%%%%%%%%%%%%  Drop formation   %%%%%%%%%%%%%%%%%%%%%%%%
%%%%%%%%%%%%%%%%%%%%%%%%%%%%%%%%%%%%%%%%%%%%%%%%%%%%%%%%%%%%%%%

\section{Drop formation: thinning and pinch-off dynamics}

We first report the qualitative thinning behavior of a bidisperse suspension compared to that of monodisperse suspensions
containing each size of particles. 
We also compare the thinning of these suspensions with the behavior of a silicone oil (AP1000)
having a shear viscosity of the same order of magnitude as a monodisperse suspension of volume fraction $\phi=40\%$.
Figures~\ref{fig:timeline}(a)-(d) shows the thinning process in the case of
(a) a viscous silicone oil without particles, 
(b) a monodisperse suspension at a volume fraction $\phi=40\%$ of 250\um particles,
(c) a bidisperse suspension at a volume fraction $\phi=40\%$ and where half of the solid volume of particles is made of 80\um particles
and the other half of 250\um particles, corresponding to $\xi = 0.5$ and $\dl/\ds=3.125$,
and (d) a monodisperse suspension with $\phi=40\%$ of 80\um particles.
These experiments illustrate that the overall thinning occurs on slightly different time scales 
between the pure liquid and the suspensions.
The initial thinning, shown in the left columns of pictures in Figure~\ref{fig:timeline},
is controlled by the effective viscosity both for monodisperse and bidisperse suspensions.
Since the silicone oil (AP1000) without particles has a shear viscosity similar to that of the suspensions,
the initial thinning dynamic is similar.
However, a significant difference due to the particles is the acceleration of the
thinning just before the pinch-off (last two images of each series in Figure~\ref{fig:timeline}).
We also observe that the presence of two particle sizes in the bidisperse suspension seems 
to play a role in the latest stages of the thinning.
Indeed, for monodisperse suspensions [figures~\ref{fig:timeline}(b) and \ref{fig:timeline}(d)]
the ligament goes from a state where it contains particles to a state where it is only made of fluid.
This change in the local composition creates a local decrease in the viscosity, 
which accelerates the pinch-off.
Such an effect was previously reported by Bonnoit \text{et al.}.\cite{bonnoit2012}
In the bidisperse case ($\xi = 0.5$, figure~\ref{fig:timeline}c), 
there is an intermediate stage where the neck only contains the smallest particles.
In summary, the initial thinning seems to be governed by the effective viscosity of the suspension,
until the neck reaches a characteristic length comparable to the particle diameter.
Under this threshold, the thinning is accelerated by the particles. 
A key difference brought by the bimodal size distribution is that the neck can contain
more than one size of particles.

\begin{figure*}
    \centering
    \includegraphics[width=0.99\textwidth]{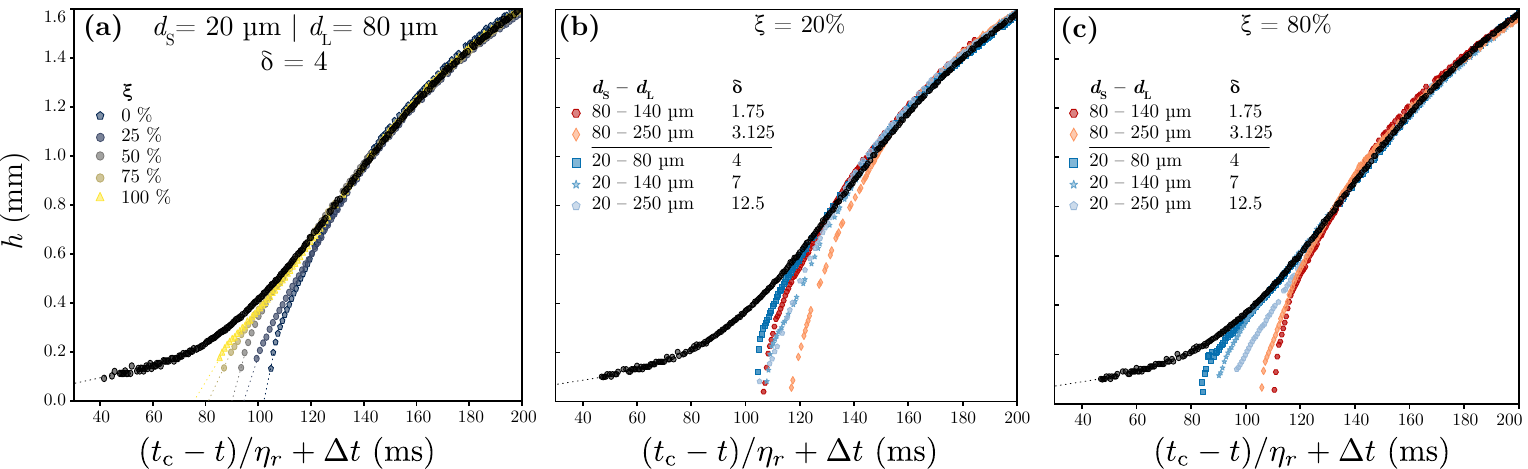}
    \caption{Rescaled thinning dynamics. 
        The experiments are the same as those presented in Figure \ref{fig:dynamics}(a)-(c).
        The time is stretched by the viscosity ratio \etar and shifted by $\Delta t$, 
        so the thinning dynamics of the suspensions overlap with that of the AP1000 oil (black circles)
        in the region before the accelerated pinch-off. 
        (a) Varying the fraction of small particles, $\xi$, 
            for a bidisperse suspension composed of particles of diameters \ds = $20\,\mu{\rm m}$ and \dl = $80\,\mu{\rm m}$ dispersed in AP100 silicone oil.
        (b) for a larger fraction of big particles ($\xi = 20\%$) and 
        (c) a larger fraction of small particles ($\xi = 80\%$)
            for different bidisperse suspensions of small and large particle diameters of \ds and \dl, 
            respectively.        
        }
    \label{fig:rescaling}
\end{figure*}

Figures~\ref{fig:dynamics}(a)-(c) report the time evolution of the minimum thickness of the neck $h$,
for suspensions involving different couples of particle sizes (\ds, \dl),
and different fractions of small particles $\xi$.
The total solid volume fraction $\phi$ is kept constant and equal to $40\%$ in all the experiments.
Note that the evolution of $h$ is plotted as a function of $\tc-t$. 
Therefore, the initial thinning is on the right side of the figures,
the pinch-off on the left, and time elapses from right to left.
For a given set of particle sizes  ($\ds = 20\um$, $\dl = 80\um$, $\delta=\dl/\ds=4$),
we recover that the final stage is slightly faster for the larger particles ($\xi=0$) than for the small ones ($\xi=1$).
In figure~\ref{fig:dynamics}(a), the yellow curve shows a suspension made of only 80\um particles, 
and the blue curve of only 20\um particles.
This acceleration becomes noticeable in the last 100-150\ms before the break-up.
The initial thinning for bidisperse suspensions is also faster than for the monodisperse suspensions
during the entire thinning process and not only the latest stages.
This observation is likely due to the fact that for a given solid volume fraction, a bidisperse suspension
has a lower viscosity than a monodisperse one.\cite{shapiro1992,pednekar2018}
The influence of the size ratios $\delta = \dl/\ds$ for a given fraction of small particles 
$\xi$ is reported in figure~\ref{fig:dynamics}(b) for $\xi=0.2$
and figure~\ref{fig:dynamics}(c) for $\xi=0.8$. 
The thinning rate increases when the size ratio $\delta$ increases.
This observation agrees with the qualitative decrease of the viscosity of the bidisperse suspension
when the size ratio increases.\cite{pednekar2018}
The acceleration of the pinch-off in the later stage also seems more significant when the large particles prevail [figure~\ref{fig:dynamics}(b)]
rather than when small particles prevail [figure~\ref{fig:dynamics}(c)].
This stronger acceleration is likely introduced by the size selection of particles 
in the ligament connecting the droplet and the nozzle.

In the following, we first consider the initial effective viscosity regime,
to relate the observed macroscopic viscosity to the viscosity of the bidisperse suspensions. 
We will then consider the accelerated thinning regime, influenced by the second size of particles.

%%%%%%%%%%%%%%%%%%%%%%%%%%%%%%%%%%%%%%%%%%%%%%%%%%%%%%%%%%%%%%%
%%%%%%%%%%%%%%%%%%% Thinning dynamics  %%%%%%%%%%%%%%%%%%%%%%%%
%%%%%%%%%%%%%%%%%%%%%%%%%%%%%%%%%%%%%%%%%%%%%%%%%%%%%%%%%%%%%%%
\section{Thinning dynamics: effective viscous-fluid regime}

We compare the thinning dynamics of the bidisperse suspensions and of the pure AP1000 silicone oil.
The viscosity of this oil, $\eta_\mathrm{AP1000}=1.28\,{\rm Pa.s}$,
is close to that of monodisperse suspensions with a volume fraction $\phi=40\%$ of particles
dispersed in the AP100 silicone oil.
Its other physical properties -- density, surface tension -- are of the same order.
The effect of the particles on the thinning is twofold.
First, far enough from the final break-up, the suspension behaves like a homogeneous fluid
with an equivalent viscosity $\eta(\phi)$.
We define the ratio of the shear viscosity of the suspension to that of AP1000 as
\begin{equation}
    \etar = \frac{\eta(\phi)}{\eta_\mathrm{AP1000}}.
    \label{eq:visc_ratio}
\end{equation}
Later, the particles accelerate the detachment of the drop compared to the equivalent fluid regime.
\cite{furbank2006}
Similarly to Bonnoit \textit{et al.}\cite{bonnoit2012} we define the time shift \dt 
between the pinch-off of the pure oil and that of the suspension.

The viscosity ratio \etar and the time shift \dt can be estimated for each suspension by adjusting
their values to find the best overlap with the AP1000 reference.\cite{bonnoit2012}
Since the typical time scale of the thinning is proportional to 
the fluid viscosity, we seek the values of \etar and \dt so that the curve
$h = f\left[(\tc-t)/\etar+\dt\right]$ for a given suspension overlaps 
as well as possible with the curve $h_\mathrm{AP1000} = f(\tc-t)$ for AP1000 oil.
This overlap is sought for in a region far away from the pinch-off,
which we define as $0.5\,\mm < h < 2\,\mm$, using an iterative bisection method.
Starting at iteration $n=0$ with $\etar^{(0)} = 1$ and $\dt^{(0)} = 0$,
we increment $\etar^{(n)}$ by one step $\Delta \eta$, 
then we compute the average time difference $\dt^{(n)}$ between
$h^{(n)} = f\left[(t_c-t)/\etar^{(n)}+\Delta t^{(n-1)}\right]$ and $h_\mathrm{AP1000} = f(t_c-t)$. 
If the mean square deviation between $h^{(n)}$ and $h_\mathrm{AP1000}$ 
increase from one iteration to the next one, 
the size of the step $\Delta \eta$ is divided by two, 
and the direction of variation of $\etar^{(n)}$ is switched.
Eventually, $\etar^{(n)}$ and $\dt^{(n)}$ converge towards \etar and \dt.
By this mean and using equation (\ref{eq:visc_ratio}),
we estimate the experimental value of the viscosity: $\etae = \etar \eta_\mathrm{AP1000}$.

Figures~\ref{fig:rescaling}(a)-(c) report the rescaling for the different bidisperse suspensions
previously reported in figures~\ref{fig:dynamics}(a)-(c).
It shows that all experiments with suspensions can be rescaled to match the initial
thinning dynamics of the pure AP1000 oil.
The deviation from the AP1000 dynamics occurs at relatively large values of $h$,
typically a few particle diameters, as observed by Bonnoit \textit{et al.}\cite{bonnoit2012}

Using this rescaling, we investigate the effect of the fraction of small particles $\xi$ 
on the pinch-off for given sizes of small and large particles, \ds and \dl, respectively (figure~\ref{fig:rescaling}a).
The monodisperse suspension of large particles ($\xi=0$, in dark blue) breaks up faster 
than the monodisperse suspensions of small particles ($\xi=1$, in yellow),
in agreement with the recent study of Ch\^ateau \textit{et al}.\cite{chateau2018}
The bidisperse suspensions also follow a monotonic trend:
as the composition shifts from the small to the large particles (when $\xi$ decreases) 
the pinch-off becomes more and more accelerated.
Besides, the more small particles are in the suspension, 
the later the thinning dynamics deviates from the reference liquid.

In Figure~\ref{fig:rescaling}(b), we compare the thinning and pinch-off behaviours 
for different diameter ratios of particle $\dl/\ds$ at a given fraction of small particles ($\xi=20\%$).
We still observe a deviation from the equivalent liquid earlier
for the bidisperse suspensions containing the biggest large particles:
typically around $h=1\mm$ for the 80--250\um suspension and $h=0.8\mm$ for the 20--250\um suspensions. 
When the large particles are smaller, this deviation is observed for smaller neck diameters, for instance, 
around 0.6\mm for the 20-80\um bidisperse suspension. 
More generally, for a given small particle diameter \ds, the larger \dl is,
the earlier the deviation from the homogeneous fluid is observed.
This behavior can be rationalized by considering that when the thickness of the neck $h$
becomes comparable to \dl,
the large particles are progressively pushed out of the neck by capillary effects.
As a result, the liquid is locally depleted in large particles.
For $\xi=20\%$, the large particles represent 80\% of the solid volume.
Therefore, the local depletion in large particles leads to a smaller particle volume fraction, 
typically dropping from $\phi=40\%$ to an estimate of $\xi\,\phi = 8\%$.
Such a change in the local volume fraction leads to
a reduction of the local viscosity by almost an order of magnitude and can explain the sudden acceleration 
of the thinning at late times.
The overall behavior is similar in the regime dominated by small particles 
and reported in Figure~\ref{fig:rescaling}(c).
However, since the large particles now only constitute 20\% of the particle volume fraction
in the bidisperse suspension,
their depletion in the neck does not lead to such a viscosity drop as drastic
as in the case with 80\% of large particles.

%%%%%%%%%%%%%%%%%%%%%%%%%%%%%%%%%%%%%%%%%%%%%%%%%%%%%%%%%%%%%%%
%%%%%     Viscosity of the bidisperse suspension    %%%%%%%%%%
%%%%%%%%%%%%%%%%%%%%%%%%%%%%%%%%%%%%%%%%%%%%%%%%%%%%%%%%%%%%%%%

\section{Viscosity of the bidisperse suspension}

\begin{figure}[t]
    \centering
    \includegraphics[width=.48\textwidth]{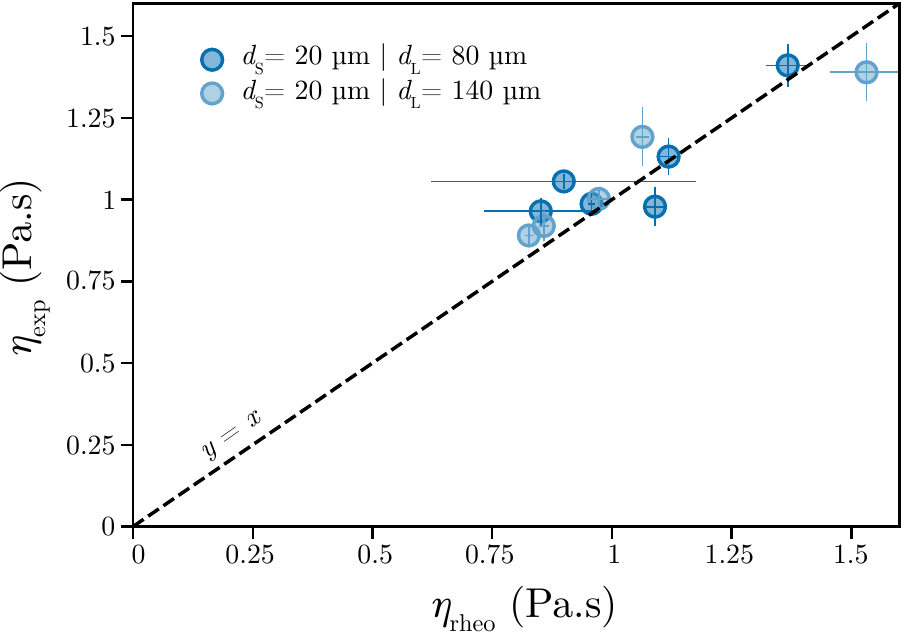}
    \caption{Comparison between the shear viscosity measured with the rheometer, $\eta_{\rm rheo}$
             and the viscosity measured from the pinch-off experiments \etae, 
             for two bidisperse suspensions and different fractions of small particles $\xi$.
             The dashed line represents $\eta_{\rm rheo}=\etae$.}
    \label{fig:rheometer}
\end{figure}

In Figures~\ref{fig:rescaling}(a)-(c), we assumed that the characteristic time scale
of the thinning process is proportional to the shear viscosity of the suspension.
Thus, we identified the fitting parameter \etar as the viscosity ratio [equation~\ref{eq:visc_ratio}].
In the most general case, the forces acting on the drop are the surface tension $\gamma$, 
the viscosity, and inertia.
Long before the pinch-off, the nozzle diameter $h_0$ is the relevant length scale,
and the thinning rate $\dot h$ is typically of order $10\,\mms$.
Therefore, the Reynolds number $Re = \rho h_0 \dot h / \eta$ for the bidisperse suspensions 
is of order $3 \times 10^{-2}$, which suggests that inertia is negligible.
In this case, the typical time scale should vary like $\eta h_0 / \gamma$,
proportionally to the viscosity.
Previous pinch-off experiments with monodisperse suspensions have shown that their Trouton's ratio
-- which compares the elongational viscosity to the shear viscosity --
is close to that of a Newtonian liquid.\cite{bonnoit2012,chateau2018}
It is therefore relevant to compare the viscosity obtained from our bidisperse pinch-off experiments,
which corresponds to an elongational flow, to the shear viscosity measured through other methods.

To compare the values obtained from the pinch-off experiments, \etae,
with the actual shear viscosity of the bidisperse suspensions,
we measured the steady-shear viscosity $\eta_\text{rheo}$ independently with a rheometer (Anton Paar MCR 92).
We used a rough parallel plate geometry of diameter 25\mm and a gap between the plates of 1\mm.
For the volume fraction considered here, $\phi=40\%$, the viscosity was found to be nearly shear-independent.
Figure~\ref{fig:rheometer} reports the shear viscosity measured with the rheometer $\eta_\text{rheo}$ 
to the viscosity estimated from the pinch-off experiments $\etae$, 
for two couples of particle sizes ($20\,\mu{\rm m }$/$80\,\mu{\rm m }$ and $20\,\mu{\rm m }$/$140\,\mu{\rm m }$).
The experimental viscosity $\eta_\text{exp}$ is obtained through equation~(\ref{eq:visc_ratio}): 
$\etae = \etar \: \eta_\text{AP1000}$.
We obtain a very good agreement between the viscosity obtained with these two different methods, 
even though the viscosity varies depending on the composition of the bidisperse suspension.
This result suggests that the viscosity of the suspension can be directly measured 
by observing the thinning of the neck.
Future studies scanning a large range of volume fractions could confirm this possibility.
Moreover, our results imply that the thinning timescale is proportional to the viscosity,
even though the thinning is not self-similar for the range of Ohnesorge numbers considered here ($2.3 < {\eta}/{\sqrt{\gamma\rho h_0}} < 4.4$).

\begin{figure*}[h]
    \centering
    \includegraphics[width=0.99\textwidth]{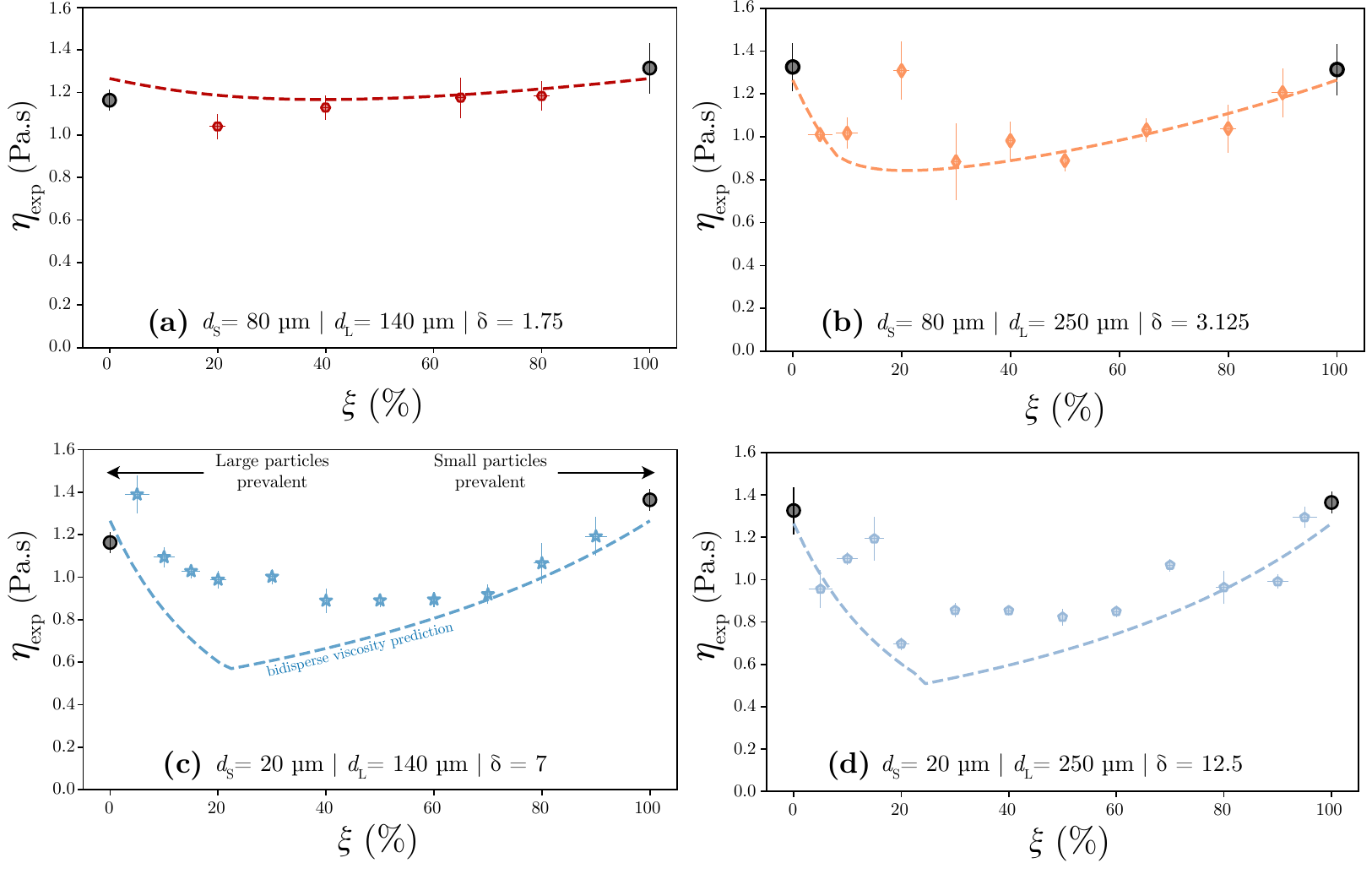}
    \caption{Effective viscosity \etae estimated from the pinch-off experiments. 
        The dashed lines represent the viscosity predicted by the model.
        (a) $\ds = 80\,\mu{\rm m}$, $\dl = 140\,\mu{\rm m}$.   
        (b) $\ds = 80\,\mu{\rm m}$, $\dl = 250\,\mu{\rm m}$.   
        (c) $\ds = 20\,\mu{\rm m}$, $\dl = 140\,\mu{\rm m}$.   
        (d) $\ds = 20\,\mu{\rm m}$, $\dl = 250\,\mu{\rm m}$.   
    }
    \label{fig:viscosity}
\end{figure*}

%%%%%%%%%%%%%%%%%%%%%%%%%%%%%%%%%%%%%%%%%%%%%%%%%%%%%%%%%%%%%%%
%%%%%%%%%%%%%%%%%% Bidisperse viscosity  %%%%%%%%%%%%%%%%%%%%%%
%%%%%%%%%%%%%%%%%%%%%%%%%%%%%%%%%%%%%%%%%%%%%%%%%%%%%%%%%%%%%%%

We now consider the variation of the viscosity observed when varying the values of $\xi$ and $\delta$.
Qualitatively, the lower viscosity of bidisperse suspensions can be explained by the more efficient
packing of spheres of different sizes.
In the limit where the large particles are much bigger than the small ones, 
the small particles can sit in the interstices 
between large particles without increasing the total volume of the packing.
Hence, the maximal solid packing fraction \phic of bidisperse suspensions is higher,
leading to a less viscous suspension for a given volume fraction $\phi$.

A variety of models have been developed to predict the value of \phic for a given particle size distribution.
In particular, Ouchiyama~and~Tanaka~\cite{ouchiyama1981} developed a model for an arbitrary polydisperse distribution.
Their approach consists of considering the local volume fraction around each particle
and then averaging it over the particle size distribution.
Gondret and Petit\cite{gondret1997} adapted this model to the specific case of a bidisperse distribution.
Their calculations leads to the maximum packing fraction
\begin{equation}
    \begin{split}
        & \phic(\delta,\xi) = \\
        & \frac{\NS \tds^3 + \NL \tdl^3}
        {(\NS/\Gamma)(\tds+1)^3 + \NL \left( (\tdl-1)^3 + \left[(\tdl+1)^3 - (\tdl-1)^3 \right]/\Gamma \right)},
    \end{split}
    \label{eq:model_phic}
\end{equation}
where \NS and \NL are the number fractions of small and large particles, respectively, and are given by:
\begin{equation}
    \NS = \frac{\xi \delta^3}{\xi \delta^3 + (1-\xi)}
    \text{\ \ \ and\ \ \ }
    \NL = \frac{1-\xi}{\xi \delta^3} \NS,
    \label{eq:model_N}
\end{equation}
and where $\tds$ and $\tdl$ are the reduced sizes given by
\begin{equation}
    \tds = \frac{\xi\delta^3 + (1-\xi) }{\xi\delta^3 + (1-\xi)\delta}
    \text{\ \ \ and\ \ \ }
    \tdl = \delta\tds.
    \label{eq:model_d}
\end{equation}
Finally, $\Gamma$ denotes the average number of particles in the vicinity of a given particle:
\begin{equation}
    \begin{split}
        \Gamma &= 1 + \frac{4}{13} (8\phi_0-1) \times \\
        & \frac{\NS(\tds +1)^2 \left( 1 - \dfrac{3}{8} \dfrac{1}{\tds+1} \right)
            + \NL (\tdl+1)^2 \left( 1 - \dfrac{3}{8} \dfrac{1}{\tdl+1} \right)}
            {\NS\,\tds^3 + \NL \left( \tdl^3 - (\tdl-1)^3\right)}
    \end{split}
    \label{eq:model_gamma}
\end{equation}
where $\phi_0$ is the maximum solid fraction in a monodisperse packing.
By fitting equation~(\ref{eq:maronpierce}) to the measurements of the viscosity for monodisperse
suspensions with various solid fractions, we measured $\phi_0 = 57.8\% \pm 0.3\%$
for all the particles used in this study.
Equation~\ref{eq:model_phic} predicts the maximal packing fraction for particles of comparable sizes
and volume fraction.
However, if the small particles are small and few enough to sit between the large ones without 
disturbing them, the maximal packing fraction is:\cite{gondret1997}
\begin{equation}
    \phic(\delta \to \infty, \xi \to 0) = \frac{\phi_0}{1-\xi}
    \label{eq:small_xi}
\end{equation}
Equation~\ref{eq:small_xi} is the upper limit for the maximal packing fraction, 
because the packing cannot be more compact than a packing where small particles are negligible in size
and in number.
Hence, if the value of \phic predicted by equation~(\ref{eq:model_phic}) is greater than $\phi_0/(1-\xi)$, 
then \phic is given by equation~(\ref{eq:small_xi}).
Once we have computed \phic, we obtain the viscosity of the bidisperse suspension
with equation~(\ref{eq:maronpierce})

Figures~\ref{fig:viscosity}(a)-(d) compare the effective viscosity measured during the pinch-off
\etae (symbols) to the theoretical value obtained using this approach (dashed line).
The grey circles represents the monodisperse cases corresponding to $\xi = 0$ (only large particles) and $\xi = 1$ (only small particles).
Our experimental data show that the viscosity of the bidisperse suspensions is systematically smaller than that of 
the monodisperse ones, and that this difference depends on the ratio of particle size $\delta=d_L/d_S$.
The larger $\delta$, the more pronounced the drop in viscosity.
The value of the viscosity when generating a droplet can drop as low as 50\% 
for the largest size ratios considered here [Figure~\ref{fig:viscosity}(d)].
In the large-particles-dominated regime, there is a very good agreement with the model 
for the couples of particle sizes that we tested ($80\,\mu{\rm m }$/$140\,\mu{\rm m }$ and $80\,\mu{\rm m }$/$250\,\mu{\rm m }$).
In the case of small size ratios [figures~\ref{fig:viscosity}(a)-(b)], 
the model predicts the viscosity during the formation of droplets on the whole range of composition.
The main deviation occurs for large size ratios [figures~\ref{fig:viscosity}(c)-(d)].
When the fraction of small particles $\xi$ is less than 50\%, the model underestimates the viscosity.
Also, the experimental variations of viscosity are more symmetrical than what the model predicts.
Although the model suggests that the viscosity should be minimum around $\xi=20\%$ for these couples of particles,
the minimum for bidisperse suspensions with a large size ratio is around 50\%.
These discrepancies are especially visible in figures~\ref{fig:viscosity}(c)-(d).
Since we have shown in Figure~\ref{fig:rheometer} that pinch-off experiments lead to viscosities similar
to those obtained with a rheometer, it seems unlikely that the higher viscosity at low $\xi$
be due to the neck geometry.
It is more likely that the problem arises from our approach to computing the viscosity since we obtain similar results with the rheometer.

In the present study, we only consider the effect of the size distribution on \phic, 
and we keep the same expression for the viscosity as a function of $\phi$ (equation~\ref{eq:maronpierce}) as for a monodisperse suspension.
The issue is that at a large size ratio ($\dl \gg \ds$), and especially when the large particles dominate 
($\xi < 50\%$), the model from Gondret and Petit predicts a steep rise in \phic. \cite{gondret1997}
This is because the small particles sit between the large ones without disturbing them.
Therefore, the bidisperse packing becomes an overlay of two separate packings that do not interact with each
other.
However, in a suspension, the small particles do interact with the large ones through lubrication films, 
whatever their size.
 
Let us consider the case $\delta \gg 1$, meaning that the small particles are very small compared to the large ones.
In the limit $\xi \to 1$, the large particles are dispersed in a suspension of small particles.
Considering that most of the viscous dissipation happens between small particles because they create stronger velocity gradients,
replacing small particles with large ones should reduce the viscous dissipation per unit volume.
As shown in Figure~\ref{fig:viscosity} this effect is well captured by the model.
In the case $\xi \to 0$, the suspension is mainly composed of large particles, 
and viscous dissipation occurs in the interstices in-between.
If we replace a small number of large particles with small ones that will sit 
in these interstices, two opposite effects play.
On one hand, the polydispersity increases \phic, so that the viscosity should decrease.
On the other hand, the small particles have little space to move between the large particles.
This confinement effect is known to increase the viscosity of monodisperse suspensions\cite{peyla2011}.
Combining these two effects possibly explains why the measured viscosity is larger
than what is predicted by the model based on the sole maximum packing fraction.
Nevertheless, this problem deserves further investigation.

%%%%%%%%%%%%%%%%%%%%%%%%%%%%%%%%%%%%%%%%%%%%%%%%%%%%%%%%%%%%%%%
%%%%%%%%%%%%%%%%%%%%% Early pinch-off  %%%%%%%%%%%%%%%%%%%%%%%%
%%%%%%%%%%%%%%%%%%%%%%%%%%%%%%%%%%%%%%%%%%%%%%%%%%%%%%%%%%%%%%%

\section{Early pinch-off}

\begin{figure}[h]
    \centering
    \includegraphics[width=.5\textwidth]{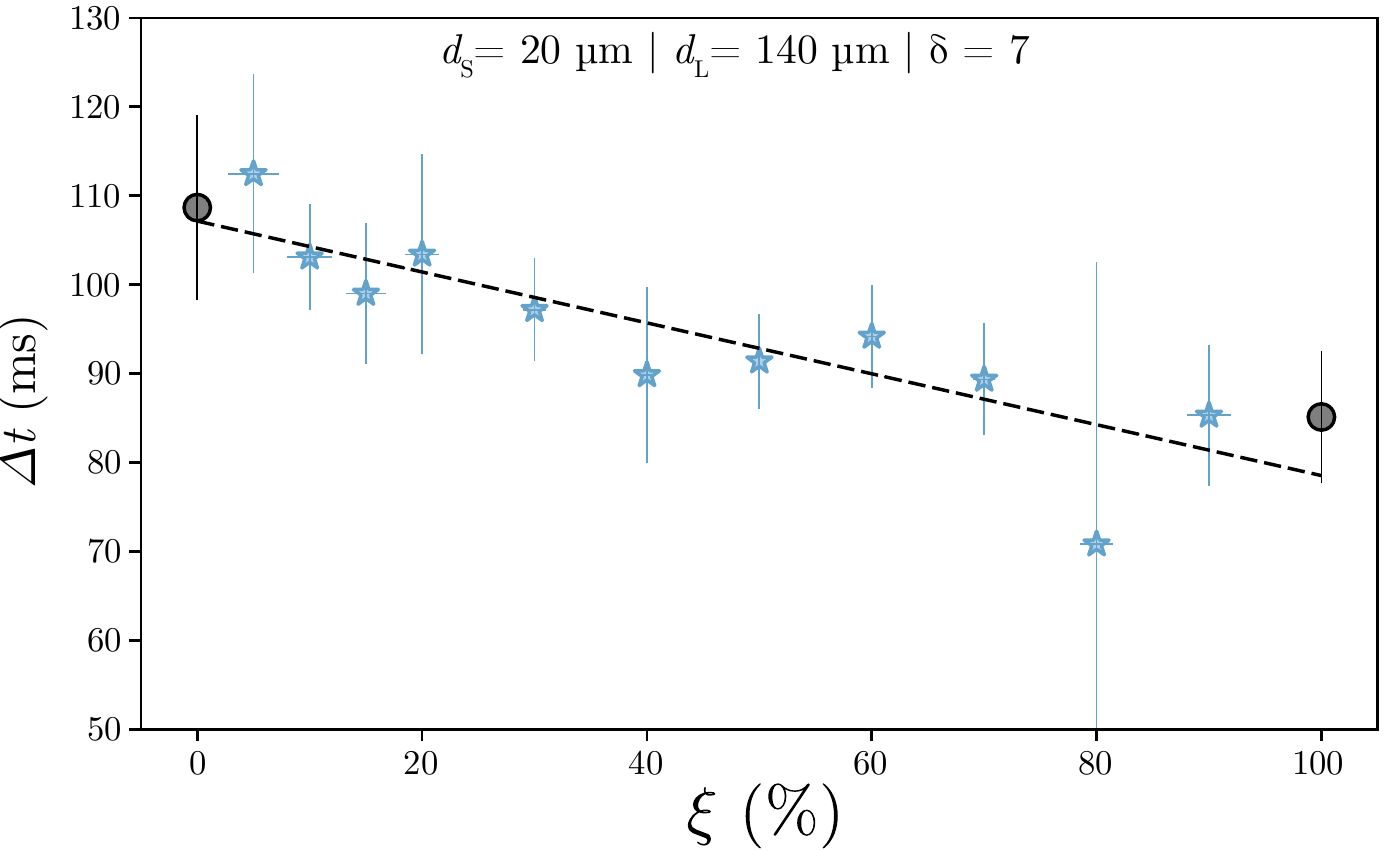}
    \caption{Time shift \dt between the pinch-off of pure AP1000 oil and that of bidisperse suspensions (\ds = $20\,\um$ and \dl = $140\,\um$), 
        as a function of the fraction of small particles $\xi$.
        The black circles represent the monodisperse cases.
        The dotted line is the best linear fit.
        }
    \label{fig:deltaT}
\end{figure}

In addition to decreasing the viscosity in the equivalent fluid regime, 
the bimodal particle size distribution acts on the discrete effects regime near the pinch-off, 
in which the thinning accelerates.
Figure~\ref{fig:deltaT} shows the evolution of the time shift \dt 
for the bidisperse suspension of 20\um and 140\um particles for various values of $\xi$.
If we first consider the monodisperse cases (black circles at $\xi=0$ and $\xi = 100\%$),
we recover that monodisperse suspensions of large particles break up earlier than those of small particles.\cite{bonnoit2012,chateau2018}
Since \dt is associated with the discrete particulate effects in the last stages before the pinch-off,
one could expect that its value is controlled by the small particles only.
However, Figure~\ref{fig:deltaT} shows that even a small amount of small particles dispersed 
amongst the large ones is enough to slightly delay the pinch-off.
Moreover, the value of \dt for intermediate bidisperse suspensions seems to vary linearly
from one monodisperse state to the other.
We observed a similar linear behaviour for the other four couples of particle sizes considered in this study.
Hence, it suggests that for a bidisperse suspension, $\dt(\xi, \ds, \dl)$ is simply the volume average of
$\dt$ over the two particles sizes:
\begin{equation}
    \dt(\xi, \ds, \dl) = \xi \dt(\ds) + (1-\xi) \dt(\dl)
    \label{eq:deltaT}
\end{equation}

Another interesting feature of the accelerated pinch-off is that it changes the thinning rate in the 
final linear regime.
In the last moments before pinch-off, the radius of a viscous liquid thread
successively follows two self-similar regimes, both linear: 
$h = 2 \veta (\tc-t)$.
First, the capillary-viscous regime, described by Eggers:~\cite{eggers1993} $\veta = 0.0304 (\gamma/\etaf)$.
Then, the inertial-viscous regime, described by Papageorgiou:~\cite{papageorgiou1995}
$\veta = 0.0709 (\gamma/\etaf)$.
For pure AP1000, we observe a linear trend (black dashed line in figure~\ref{fig:dynamics}(a)), 
with a thinning rate \veta of the same order of magnitude as predicted by Eggers and Papageorgiou.
Interestingly, for the suspension shown in figure~\ref{fig:quali}(c) the thinning rate in the linear regime
is twenty times larger than the value predicted by Eggers,\cite{eggers1993}
and ten times larger than that predicted by Papageorgiou.\cite{papageorgiou1995}
This result suggests that in the last instants before pinch-off, although the viscous thread is devoid of
particles, it is still subject to their influence.
Nevertheless, these aspects of accelerated pinch-off deserve a further study, 
notably considering the local variations of $\phi$ and $\xi$ at the neck.

%%%%%%%%%%%%%%%%%%%%%%%%%%%%%%%%%%%%%%%%%%%%%%%%%%%%%%%%%%%%%%%
%%%%%%%%%%%%%%%%%%%%%%% Conclusion  %%%%%%%%%%%%%%%%%%%%%%%%%%
%%%%%%%%%%%%%%%%%%%%%%%%%%%%%%%%%%%%%%%%%%%%%%%%%%%%%%%%%%%%%%%

\section{Conclusion}

In this study, we have characterized the thinning and pinch-off of drops of suspensions 
with a bimodal particle size distribution. 
We found that the size of the particles, as well as the relative fraction of each size, influence
both the equivalent fluid regime, where suspensions can be considered as liquids with a larger effective viscosity, and the discrete regime, where particles accelerate the thinning and the break-up.
We demonstrated that the time scale associated with the thinning was proportional to the suspension viscosity.
We were then able to rescale the thinning dynamic of each suspension 
to that of a reference liquid of known viscosity. 
This method enables a direct measurement of the viscosity of bidisperse suspensions.
Our experiments also reveal how the composition of bidisperse suspensions influences the late discrete regime.
This regime is characterized by a time shift \dt between the pinch-off of the suspension
and that of pure silicone oil of comparable shear viscosity.
Our study suggests that for a bidisperse suspension, the value of \dt varies linearly 
from its value for the small particles to its value for the large particles.
Also, although its scope is limited to bidisperse suspensions, this study aims towards polydisperse suspensions, a relevant system for industrial processes.
Recent numerical simulations have shown that the rheology of a polydisperse suspension could be linked
to that of one with a statistically equivalent bidisperse distribution.\cite{pednekar2018}
Moreover, we believe that proper implementation of the Ouchiyama-Tanaka model\cite{ouchiyama1981} would
extend the prediction of the viscosity to suspensions with an arbitrary size distribution.
Eventually, controlling the viscosity drop induced by the polydispersity could help 
improving the efficiency of printing methods.

\section*{Conflicts of interest}
There are no conflicts to declare.

\section*{Acknowledgements}
We gratefully acknowledge the discussions with Philippe Gondret regarding the viscosity of bidisperse suspensions. This material is based upon work supported by the National Science Foundation under NSF CAREER Program Award CBET Grant No. 1944844 and by the ACS Petroleum Research Fund Grant No. 60108-DNI9.

%%%END OF MAIN TEXT%%%

\balance

\bibliography{pinch-off_bidisperse} 
\bibliographystyle{unsrt}

\end{document}